# Quantitative Comparison of Statistical Methods for Analyzing Human Metabolomics Data

Brian Claggett,*[1] Joseph Antonelli,*[1,2] Mir Henglin,[1] Jeramie D. Watrous,[3] Kim A. Lehmann,[3] Gavin Ovsak,[1] Gabriel Musso,[1] Andrew Correia,[1] Sivani Jonnalagadda,[1] Olga V. Demler,[4] Ramachandran S. Vasan,[5,6] Martin G. Larson,[6,7] Mohit Jain,*[3] Susan Cheng*[1,6]

*Authors contributed equally

[1]Cardiovascular Division, Brigham and Women's Hospital, Harvard Medical School, Boston, MA

[2]Department of Biostatistics, Harvard T.H. Chan School of Public Health, Boston, MA

[3]Departments of Medicine & Pharmacology, University of California San Diego, La Jolla, CA

[4]Preventive Medicine, Brigham and Women's Hospital, Harvard Medical School, Boston, MA

[5]Preventive Medicine, Department of Medicine, Boston University Medical Center, Boston, MA

[6]Framingham Heart Study, Framingham, MA

[7]Biostatistics Department, School of Public Health, Boston University, Boston, MA

**Correspondence:** Susan Cheng, MD, MMSc, MPH, Brigham and Women's Hospital, Harvard Medical School, scheng@rics.bwh.harvard.edu, and Mohit Jain, MD, PhD, University of California, San Diego, mjain@ucsd.edu.

**ABSTRACT**

**Background.** Emerging technologies now allow for mass spectrometry based profiling of up to thousands of small molecule metabolites ('metabolomics') in an increasing number of biosamples. While offering great promise for revealing insight into the pathogenesis of human disease, standard approaches have yet to be established for statistically analyzing increasingly complex, high-dimensional human metabolomics data in relation to clinical phenotypes including disease outcomes. To determine optimal statistical approaches for metabolomics analysis, we sought to formally compare traditional statistical as well as newer statistical learning methods across a range of metabolomics dataset types.

**Results.** In simulated and experimental metabolomics data derived from large population-based human cohorts, we observed that with an increasing number of study subjects, univariate compared to multivariate methods resulted in a higher false discovery rate due to substantial correlations among metabolites. In scenarios wherein the number of assayed metabolites increases, as in the application of nontargeted versus targeted metabolomics measures, multivariate methods performed especially favorably across a range of statistical operating characteristics. In nontargeted metabolomics datasets that included thousands of metabolite measures, sparse multivariate models demonstrated greater selectivity and lower potential for spurious relationships.

**Conclusion.** When the number of metabolites was similar to or exceeded the number of study subjects, as is common with nontargeted metabolomics analysis of relatively small sized cohorts, sparse multivariate models exhibited the most robust statistical power with more consistent results. These findings have important implications for the analysis of metabolomics studies of human disease.

**BACKGROUND**

Mass spectrometry based measurements of small molecule metabolites, also known as metabolomics, has emerged as a powerful tool for phenotyping biochemical variation in health and disease across organisms. Accordingly, there has been a rapidly growing interest in applying metabolomics to clinical studies of human disease traits.[1,2] Metabolomics technologies have recently advanced from the measure of approximately 200 distinct small molecules in a 'targeted' fashion to reproducible quantification of up to several thousand small molecules using 'nontargeted' approaches. Such technical advances have emphasized the need to determine optimal methods for the statistical analysis of high-dimensional metabolomics data. Robust statistical methods are particularly needed to examine associations of metabolites detected in peripheral blood circulation with disease traits in humans; in this context, false discovery remains a key concern for clinical biomarker studies.[3-5] The statistical analysis challenges posed by human metabolomics data arise from multiple sources. For instance, metabolomics data collected from a given biospecimen represents metabolite variation at a particular point in time and in a particular context: whereas a portion of the variability reflects the relatively stable components of the organismal metabolome, another component reflects the dynamic portion of the metabolome that varies substantially over time and in response to a number of exposures. Such mixed structure can lead to a high degree of variation for a given metabolite level across individuals. Additionally, due to common pathways of enzymatic production or exposures of origin, metabolites can demonstrate a high degree of inter-correlation, and this inter-correlation may vary between individuals or subgroups depending on disease state, exposures, or other factors.

Initial clinical studies involving targeted metabolomics approaches have used relatively conservative statistical approaches to analyzing up to 200 variables, such as Bonferroni correction of multiple t-tests or the false discovery rate (FDR).[6] Additionally methods of accounting for multiple hypothesis testing have similarly assigned more or less conservative thresholds for defining statistical significance. In the absence of considering inter-correlations between individual metabolites at the outset, data analyses will tend to favor identifying metabolites from a singular biological pathway, with secondary or tertiary associations (potentially representing important orthogonal pathways) being forced to reach lower levels of statistical significance based on rank ordering alone. For this reason, traditional statistical approaches are believed to offer limited sensitivity for high-dimensional metabolomics analyses. Thus, several alternative methods have been proposed to more effectively select metabolites associated with a given outcome.[7-11] These methods have begun to surface from analyses of other molecular phenomics datasets,[9,10,12,13] although they may differ in structure relative to metabolomics datasets. Each statistical method has intrinsic strengths and weaknesses, and the extent to which they may be more or less suited for a given metabolomics analysis is not known but is likely to depend on number of metabolites assayed, sample size, and frequency or type of clinical outcome. Therefore, we sought to formally test currently available statistical methods across a range of dataset types. By simulating clinical studies to test different outcomes-based hypotheses and validating findings using real metabolomics data, we aimed to assess the suitability of statistical methods for the analysis of metabolomics data across a range of clinical data settings.

## METHODS

**Development of Simulated Metabolomics Dataset**

We developed a series of simulated metabolomics datasets based on the characteristic data features seen in both experimental and human studies (small case-control as well as large cohort) using both targeted and nontargeted mass spectrometry methods. In particular, we designed the datasets to include the range of structural characteristics typically observed in human plasma metabolomics datasets.

Structural features with respect to outcomes included: (i) binary outcomes in small-sized studies of up to 200 individuals with an outcome frequency of 50%, representing case-control studies with a 1:1 case/control ratio; (ii) binary outcomes in large-sized studies with an outcome frequency of 20% in cohorts with >200 individuals, representing larger observational cohort studies; (iii) continuous outcomes measured in all patients. Structural features with respect to exposures included: (i) number of metabolites ranging from 200 (as is typical of a targeted method) to 2000 (representative of an nontargeted method); (ii) metabolite values following a normal distribution, which is similar to what is commonly observed after a logarithmic or other transformation of the data; (iii) general positive correlation between metabolites, with pairwise correlations randomly distributed around a mean of +0.40; (iv) clustering within the data such that large groups of metabolites are highly correlated with each other representing biological pathways; and (iv) the number of "true positive" metabolites independently associated with the clinical outcome set to 10, with varying effect sizes. A summary of the data structures is provided in **Table 1** and an example correlation matrix induced by our simulation design is shown in **Figure 1**.

**Statistical Approaches for Analyzing Metabolomics Data**

For comparison of analyses, we applied the following statistical methods to the simulated metabolomics datasets: (1) univariate analyses with multiple testing correction using the Bonferroni or false discovery rate (FDR) approach;[6] (2) principal component regression (PCR);[14,15] (3) sparse partial least squares (SPLS);[8,10] (4) sparse partial least squares discriminant analysis (SPLSDA);[8,9] (5) random forests;[7] and (6) least absolute shrinkage and selection operator (LASSO).[11] Univariate analyses with multiple testing correction, including Bonferroni correction or the Benjamini-Hochberg correction for FDR, have been applied in a variety of predominantly targeted metabolomics studies previously.[16-18] The PCR approach, also applied in prior metabolomics studies, reduces the dimension of the total number of metabolite variables. PCR first reduces the dimensionality of the metabolite data, then uses the selected principal components in a regression model to predict the clinical outcome variable. Finally, variable importance measures are derived by reallocating the estimated regression coefficients to the metabolites that contributed to each of the chosen principal components. The PLS regression method maximizes the covariance between a matrix of metabolite variables and the outcome variable, where the outcome is typically a continuous variable; for categorical outcome variables, a variant called PLS discriminant analysis [PLSDA][19] may be applied. Either PLS or PLSDA serve to decompose metabolite and outcomes data into latent structures and aim to maximize the covariance between these latent structures. The random forests method employs a non-parametric ensemble approach to predicting an outcome from metabolomics data by identifying presumably non-linear patterns that may account for metabolite variation in relation to a particular outcome.[7] PCR, PLS, PLSDA, and random forests all suffer from a similar problem when trying to identify important metabolites: While they can rank order the metabolites in terms of importance, there is no obviously principled way to select a cutoff for which metabolites are 'significantly' associated with the outcome. There exist ad hoc approaches to performing variable selection in some of these contexts;[20,21] however, there is no consensus on the appropriate manner for selecting important metabolites. Naïve approaches such as simply taking the top K covariates to be significant can be applied, but their properties are not well understood and their performance will vary greatly across data sets depending on the true number of significant metabolites present. One way to overcome this issue is to use models that induce sparsity in their respective coefficients. Sparsity refers to the assumption that, adjusting for all measured metabolites, the number of metabolites that are truly associated with the clinical outcome (true positives) is far smaller than the overall number of metabolites. The most popular such approach in the field of statistics is LASSO,[11] a method that regresses a given outcome on all metabolite variables simultaneously and achieves parsimonious variable selection by applying a penalty to the magnitude of the regression coefficients. Because many statistical methods are unable to simultaneously model a number of metabolites which exceeds the number of study subjects, the assumption of sparsity allows for many more traditional methods to be extended to such high-dimensional data. Notably, sparse extensions of approaches such as PLS and PLSDA, exist[7-10] and are useful for their application in metabolomics. Sparse extensions of these methods provide automatic variable selection, which solves the aforementioned issue that these methods only allow for variable importance ranking. The FDR approach was implemented using the Benjamini-Hochberg procedure with a false discovery rate of 0.1. Parameters of multivariate approaches such as LASSO, SPLS, and SPLS-DA were selected using cross-validation. Variables were ranked based on the magnitude of the absolute value of the relevant regression coefficients for LASSO, SPLS, and PCR, while variable importance measures were used for random forests

It is also important to clarify the distinction between variable selection and significance testing. Methods such as the LASSO or other sparse models do not perform significance testing in the traditional sense of controlling type I error or testing hypotheses. Rather, they simply identify a set of metabolites that are relatively important for predicting a given outcome. Thus, herein, we will compare approaches aimed at identifying metabolites of greatest interest, in relation to a given outcome, wherein some approaches involve traditional hypothesis testing while others involve simply variable selection.

To compare the performance between statistical methods in this regard, we evaluated the following metrics: (i) probability of identifying a true positive metabolite through variable selection/significance testing as function of the true effect size (among those methods which allow for such identification); (ii) probability of identifying a true positive metabolite as a "top 10" metabolite as a function of effect size; (iii) average number of false positive metabolites identified by variable selection/significance testing; (iv) positive predictive value (PPV), the probability that a metabolite identified through variable selection/significance testing is truly related to the clinical outcome; (v) negative predictive value (NPV), the probability that a metabolite not identified is truly unrelated to the clinical outcome. These metrics were evaluated separately for continuous and binary outcomes, with all analyses performed using *R*v3.2.3 (R Development Core Team, Vienna, Austria). To visualize the relatedness of metabolites, Spearman correlation coefficients were estimated for all pairs of metabolites and correlations above ≥0.65 were isolated, with clusters of these correlations visualized using a D3 visualization framework (https://github.com/d3/d3-force); within this visualization, metabolites associated with sex and age via the Bonferroni, FDR, SPLS, and LASSO methods were highlighted.

**Experimental Human Metabolomics Data**

As part of the community-based Framingham Heart Study, the offspring cohort participants underwent a standardized evaluation that included fasting blood sample collection at their eighth examination in 2002-2005, as previously described.[22] All participants provided informed consent and all protocols were approved by the institutional review boards at Boston University Medical Center, Brigham and Women's Hospital, and the University of California, San Diego. LC-MS based metabolomics analysis was performed on all available N=2895 plasma samples, according to previously described protocols.[23] In brief, plasma samples were prepared and analyzed using a Thermo Vanquish UPLC coupled to a high resolution Thermo QExactive orbitrap mass spectrometer. Metabolites were isolated from plasma using protein precipitation with organic solvent followed by solid phase extraction. Extracted metabolites underwent chromatographic separation using reverse phase chromatography whereby samples were loaded onto a Phenomenex Kinetex C18 (1.7um, 2.1x100mm) column and eluted using a 7 minute linear gradient starting with water : acetonitrile : acetic acid (70:30:0.1) and ending with acetonitrile : isopropanol : acetic acid (50:50:0.02). LC was coupled to a high resolution Orbitrap mass analyzer with electrospray ionization operating in negative ion mode, with full scan data acquisition across a mass range of 225 to 650 m/z. Thermo .raw data files were converted to 32-bit centroid .mzXML using Msconvert (Proteowizard software suite), and resulting .mzXML files were analyzed using Mzmine 2.21, as described.[23] To eliminate redundant and non-metabolic chromatographic features, we applied the following filters: naturally occurring $^{13}$C isotopes were consolidated under the monoisotopic peak; common adducts (i.e. $H^+$, $Na^+$, $NH_4^+$, and $K^+$ for positive mode and $H^-$, $Cl^-$, and acetate for negative mode) were consolidated with the most abundant species being reported; multiple charge states were consolidated with the singly charged state with the most abundant being reported; and, common ESI in-source fragments (e.g. loss of water) were removed. In addition, all chromatographic features present in a sample blank subjected to the entire sample preparation protocol (with the exception of water being used instead of plasma) were removed. Finally, all remaining chromatographic features were manually inspected for quality in peak shape, retention time consistency, and signal to noise ratio, with features exhibiting sub-par characteristics subsequently removed. Known compounds that are typically observed in human plasma when applying this method are listed in **Table S1**.

From plasma collected from N=2895 participants, a total of 1933 distinct metabolite species were measured with a non-missing value recorded for every participant. We log transformed and standardized all metabolites to have mean 0 and standard deviation 1 due to the expectedly right-skewed nature of the data. Using the same statistical analytical methods described above, we conducted analyses to identify distinct metabolites demonstrating significant associations with age and sex. These phenotypes were specifically selected given both are basic factors available in almost all biomarker analyses, and they allow for analysis of a continuous and binary outcome, respectively.

**RESULTS**

**Statistical Analyses of Simulated Metabolomics Data**

Metabolomics studies of human samples can vary substantially by sample size, the number of metabolites assayed, and the type and frequency of a clinical outcome of interest, with each of these factors potentially influencing statistical analysis results. To evaluate statistical methods for handling of a variety of datasets, metabolomics data were simulated for clinical studies of varying number of study subjects, number of metabolites, and outcome type (continuous vs binary). A total of six traditional statistical (Bonferroni, FDR, and PCR) and statistical learning (LASSO, SPLS, and random forest) methods were used to analyze 1000 simulated metabolomics datasets (**Figures 2-3**), with each evaluated for the likelihood of a metabolite being correctly identified as one of the top 10 most important metabolites with respect to a given outcome. For a simulated continuous outcome (**Figure 2**), all approaches performed similarly well, with the exception of scenarios with a large number of metabolites (M=2000) or a small number of subjects (N=200). At these extremes, multivariate approaches based on sparsity, LASSO and SPLS, were found to outperform univariate approaches. In the case of a binary outcome (**Figure 3**), optimal statistical methods were less apparent. Univariate approaches based on the linear model performed slightly better than multivariate approaches with small sample sizes. As the sample size increased, results approximated those observed in the continuous case, where sparse multivariate methods such as SPLSDA outperformed the other approaches (**Figure 3**).

An equally important aspect of a statistical procedure is the identification of important metabolites via variable selection or significance testing. Variable selection is not generally possible with PCR and random forest analyses, precluding assessment of these approaches for prioritizing individual metabolites. In either the continuous or binary settings, univariate approaches performed worse as the number of study participants increased (**Figure 2-3**). While counterintuitive given that statistical performance in general is enhanced with sample size, due to the frequently correlated nature of metabolomics data, false positives increased greatly with univariate methods as non-significant variables are identified due to their correlation with significant variables (**Figure 2-3**). This result contributed to poor positive predictive value and reduced specificity for any of these approaches, both of which are important concerns for clinically relevant biomarker discovery. Multivariate approaches, by contrast, do not suffer this same drawback as their performance improves as the sample size increases (**Figure 2-3**). In the case of a continuous outcome, both LASSO and SPLS methods performed remarkably well, with SPLS slightly outperforming LASSO in terms of positive predictive value, negative predictive value, and number of false positives. Again, binary outcomes differed from statistical analysis of continuous outcomes, due to different performance for the respective estimators at different sample sizes. In small sample sizes, univariate procedures with a multiplicity correction had the best positive predictive value among all estimators (**Figure 3**). As the sample size increased to 1000 or 5000, the multivariate approaches again outperformed the univariate procedures as both LASSO and SPLSDA obtained the highest positive predictive value, negative predictive value, and the fewest number of false positives identified. Interestingly, for SPLSDA, the positive predictive value decreased from N=1000 to N=5000 as the number of false positives increased, although this was likely due to sensitivity of tuning parameter selection, which is required for the application of sparse methods.

Although we aggregated over all 10 significant metabolites to calculate measures of method performance, differences in operating characteristics such as power likely correspond with smaller effect sizes. While most methods will identify associations with large magnitudes of effect, potentially important discrepancies can become apparent for associations with smaller effect sizes. Thus, we also examined variation in power across a range of effect sizes for different statistical approaches and these results were concordant with those of aggregated data (**Figures S1-S2**).

We also constructed simulations that involved scenarios that included negative as well as positive between-metabolite correlations, in addition to pairs of highly inter-correlated metabolites representing molecular markers putatively derived from the same biological pathway. We observed that the results of these additional simulations were very similar to those produced by the primary simulations and reported herein (**Figure S3**-**S6**), suggesting that our overall findings from the simulated data are relatively consistent across variations in simulated data structure. Collectively, these findings suggest the value of multivariate approaches for identifying metabolite markers that are associated with clinical traits.

**Statistical Analyses of Experimentally Derived Metabolomics Data**

Although the above reported results put forth a statistical framework for considering analysis of clinical metabolomics based on analyses of simulated data, we sought to compare our findings with those using actual "real world" experimentally derived metabolomics data. For these analyses, we used a nontargeted metabolomics based panel of 1933 metabolites measured across 2895 individuals (see *Methods*). We restricted attention to the methods that would easily allow for individual variable (i.e. metabolite) importance selection in this dataset, which precluded random forests and PCR from entering into the analysis. Analyses using the 3 main statistical approaches (FDR, LASSO, and SPLS) revealed overlap (**Figure 4**) for only a minority of the total detected associations between metabolites and either a continuous variable (age) or a binary variable (sex). We excluded from the Venn diagram the results from the Bonferroni correction, given it produces a subset of the same metabolites chosen using the less conservative FDR correction. We applied a false discovery rate of 0.1, which suggests that 10% of the metabolites on average should be false discoveries. For both outcomes, use of the FDR resulted in a large number of statistically significant results with >50% of all assayed metabolites (1281/1933 for age, 1312/1933 for sex) reaching threshold, suggesting that an FDR correction of 0.1 was in fact detecting nearly all of the signals. The approaches rooted in sparsity, however, obtained solutions with far fewer metabolite "hits". In both cases, the LASSO analysis resulted in far fewer metabolites than an FDR correction (206 for sex, and 378 for age). By contrast, SPLS provides a solution with far fewer metabolite hits than either LASSO or an FDR correction (93 for sex, and 37 for age). In the case of both age and sex, SPLS did not identify any new metabolites beyond those found in the LASSO or FDR subsets. We found in this study that when implementing cross validation to estimate tuning parameters of SPLS and SPLSDA, the cross validation curve is relatively flat, a previously encountered issue.[8] This suggests that different levels of sparsity were equally supported by the data, and we chose to use the most sparse option to identify the most important metabolites. SPLS is a relatively new approach for which these issues have not been well addressed, and this differs for the LASSO approach wherein estimating the tuning parameter is straighforward using *glmnet* in *R*.[24] We repeated all analyses in the Framingham Heart Study dataset while including an indicator variable for specimen batch in all models. As shown in **Figures S7**, results of these analyses were similar to those in the original analyses

To provide further context for these results, we used a basic network analysis to visualize correlations between metabolite measures in the Framingham cohort and compared the relative location of metabolites identified in association with age or sex by the univariate and multivariate methods evaluated (**Figure 5**). We observed that univariate approaches tend to identify highly inter-correlated metabolites, whereas multivariate approaches exhibit a more parsimonious as well as broader selection of metabolites.

**Results from Both Simulated and Experimentally Derived Data**

The multiple statistical analysis approaches, when applied to both the simulated and the experimentally derived data, produced relatively comparable results with respect to very large numbers of metabolite markers identified by univariate compared to multivariate methods. Of the multivariate methods evaluated, the LASSO approach appeared to perform slightly better than the SPLS approach across the different types of simulated data structures and especially those involving larger numbers of metabolites (**Figures 2-3**). In comparison, the SPLS approach appeared to be more selective when applied to the experimentally derived data set (**Figure 4**). Given that selectivity alone is not necessarily a measure of true association, these results together suggest that results of either SPLS or LASSO would be reasonable to consider in a clinical study, particularly given that the metabolites selected by SPLS were identified in association with either the continuous or binary outcomes overlapped with those identified by the univariate or alternate multivariate methods.

**DISCUSSION**

Through an extensive simulation study, we investigated the relative merits of traditional statistical and statistical learning approaches for the analysis of human metabolite data. Using a data structure based on real-world metabolite data with varying sample size, metabolite number, and outcome measures, our results offer a framework for considering optimal statistical approaches for a given study. We found that penalized approaches favoring sparsity led to substantially improved inference for a wide range of scenarios. Both the LASSO and SPLS (SPLSDA for binary outcomes) procedures provided reasonable results in all simulation scenarios studied, identifying important metabolites without suffering from large numbers of false positives. The only scenario wherein univariate procedures would be most reliable was when the sample size was small and the outcome was binary. With a binary outcome, there is relatively little information available to identify associations among a very large number of metabolites; thus, approaches that attempt to model all metabolites at once do not perform as well with smaller sample sizes. Interestingly, we observed the counterintuitive phenomenon that univariate procedures perform worse at identifying significant metabolites as the sample size grows. This appeared due to the correlation structure present in the data, which leads to a large number of false positives, and presents a finding with important implications for future analyses of nontargeted metabolomics data.

Much of the current human metabolomics literature has relied on univariate approaches with a Bonferroni or FDR correction procedure, PCA, or PLSDA without penalization, in the absence of any formal evaluation of optimal statistical methods. While such approaches have proved useful in some respects for analyzing metabolite data, our findings indicate that these approaches may suffer major drawbacks in certain situations. Univariate approaches as discussed above can lead to misleading results when the data are inter-correlated, as is nearly always the case in metabolomics studies given common biochemical and biological origins. Other approaches such as PCA or PCR do not provide measures of statistical significance, and only provide ad-hoc measures of variable importance. While metabolomics data may offer some unique challenges, including scale in metabolite levels and missingness across a population, as well as biologically driven inter-correlations, related molecular phenomics fields have similarly suggested that newer statistical approaches may be of great value in identifying statistical significance and prioritization of variables for biological follow up.[13,25-29] Our results suggest that approaches relying on sparsity to perform variable selection lead to quite good performance with respect to all the metrics examined and represent a path forward for future analyses. In particular, when the number of metabolites was similar to or exceeded the number of study subjects, sparse multivariate models exhibited robust statistical power with consistent results as expected given the design of methods that prioritize sparsity. Thus, nontargeted metabolomics analyses of relatively small sized cohorts are most likely to benefit from using sparse multivariate models in attempts to identify metabolites associated with a given outcome.

There is in metabolomics a strong interest in pathway analyses that might offer some insights regarding the biological mechanisms underlying statistical associations observed between metabolites and a given outcome. These approaches remain in development, given ongoing challenges related to compound identification and limited knowledge regarding putative biological pathways relevant to novel metabolites.[30] Thus, we elected to use a relatively basic network analysis to visualize inter-metabolite associations and further examine our main findings in this context. We found that univariate approaches tend to identify highly inter-correlated metabolites, whereas multivariate approaches exhibit a more parsimonious as well as broader selection of metabolites. In effect, these findings also support the notion that multivariate compared to univariate approaches tend to select metabolites representing putative distinct and likely more orthogonal biological pathways of potential importance and interest in relation to a given outcome.

In our study, we found that multivariate approaches that assume some level of sparsity by design (i.e. some metabolites have a very small effect on the outcome) perform with the greatest efficiency for identifying important metabolites. Importantly, this conclusion is based on settings in which the relationship between a metabolite and outcome is linear. In the setting of nonlinearity, it is likely that random forest or other machine learning based approaches that allow for highly nonlinear relationships may be preferable, although this would require more formal evaluation than provided herein. The value of sparse multivariate analysis may be due to several potential reasons, including the large amount of correlation between metabolites that requires approaches to examine a metabolite conditional on the other metabolites. In addition, the fact that many metabolites indeed have little to no association with an outcome of interest favors approaches that enforce sparsity. With these results in mind, we can provide recommendations for future analysis of high dimensional metabolite data. For larger (>1000) sample sizes, multivariate approaches based on sparsity provide a very reasonable strategy to identifying important metabolites. In small sample sizes (<200), particularly for binary outcomes, there is no clear cut 'best' method and the merits of each method will depend heavily on the structure of the data. In these cases, utilizing more than one analysis tool in conjunction could help identify key covariates. Importantly, the goal of the study should be taken into account before selecting a statistical approach. If hypothesis generating discovery of potentially important metabolites is of the utmost importance, and there is little penalty for false positives, then one can use all the proposed analyses, even those methods that tend to produce a large number of false positive results. If, however, false positives are very undesirable, then we recommend approaches such as LASSO or SPLS that impose sparsity into the model and tend to eliminate presumably less relevant metabolites.

Our findings can be used to guide the design of future studies, particularly those for which investigators may be interested in estimating a minimum amount of statistical power to detect an association of interest. If pre-existing knowledge of the distribution of metabolite values and their correlation structure is available, either through preliminary or previously analyzed data, then simulation studies similar to those described in this manuscript could be performed. Investigators can use an observed correlation structure to simulate a dataset of metabolite measures and then simulate outcomes given a known range of effect sizes, from which power to detect these effects can be estimated for different statistical approaches.

There are several limitations of the study that merit consideration. The primary findings were based on simulated data, albeit data constructed based on the known structure of an existing high-dimensional data set derived from actual values in a human cohort. As such, our results may have been influenced by the nature of the underlying artificially created data structure. For this reason, we conducted parallel analyses in a de novo real-world dataset of metabolomics measures performed in a community-based cohort, and observed results that were largely consistent with those of the simulated data analyses. The observed substantial difference in performance between traditional statistical and statistical learning methods may well have emerged from the difference between univariate and multivariate methods. Accordingly, investigators have suggested that in situations where intercorrelations among predictor variables are expected, a permutation-based FDR approach to univariate analyses should be considered.[31] The extent to which FDR with permutation, or similar variants of univariate analyses, could effectively accommodate correlations and produce different results remains unclear and a subject of ongoing research.[32] It also should be emphasized that validity of results produced by any statistical model depends not only on model characteristics but also on data quality, which relies on mass spectrometry methods for correctly and consistently identifying metabolites from typical background artefact.[33] Thus, all statistical analyses are at risk for results of association analyses to be biased to the null due to technical mis-interpretations of noise for signal. It is important to note that this manuscript is focused on evaluating statistical approaches aimed at identifying metabolites with potential true associations with a given outcome, based on the goal of discovering possible underlying mechanisms. By contrast, another distinct goal in clinical metabolomics research is to identify metabolites most important for the purpose of predicting a given outcome; to this end, univariate approaches may be equivalent or superior to multivariate approaches and this is an area that warrants future investigation.

In summary, our findings further indicate that statistical learning approaches aimed at modeling a high-dimensional set of metabolites and their associations with a given outcome warrant more attention in the literature. Taken together, our results suggest that metabolomic analyses should shift towards use of multivariate approaches for identifying distinct markers associated with clinical traits. Univariate approaches, while simple to use, will identify large numbers of false positives when the metabolites are highly correlated with each other – a problem ubiquitous in metabolomics research. If interest lies solely in finding large, biologic pathways instead of causal markers (i.e. hypothesis-generating analyses), then univariate approaches may still be useful.

When compared to traditional and frequently employed univariate approaches, statistical learning methods (such as LASSO or SPLS) offer effective and easy to implement options for handling high-dimensional, correlated data of the nature that is commonly seen in metabolomics. In fact, these approaches may well outperform many of the conventionally used methods across a wide variety of scenarios encountered in human metabolomics studies.

**CONCLUSIONS**

We examined the results of applying both traditional statistical and statistical learning methods across a range of metabolomics datasets. We observed that when the number of metabolites was similar to or exceeded the number of study subjects, as is common with nontargeted metabolomics performed in small sized cohorts, sparse multivariate models demonstrated the most consistent results and the most statistical power. These findings have important implications for the analysis of metabolomics studies of human disease.

**LIST OF ABBREVIATIONS**

FDR: false discovery rate

LC-MS: liquid chromatography mass spectrometry

LASSO: least absolute shrinkage and selection operator

NPV: negative predictive value

PCR: principal component regression

PLS: partial least squares

PLSA: partial least squares discriminant analysis

PPV: positive predictive value

SPLS: sparse partial least squares

SPLSDA: sparse partial least squares discriminant analysis

UPLC: ultra performance liquid chromatrography

**DECLARATIONS**

**Ethics approval and consent to participate.** All participants provided informed consent and all protocols were approved by the institutional review boards at Boston University Medical Center, the Brigham and Women's Hospital, and University of California San Diego.

**Consent for publication.** Not applicable.

**Availability of data and material.** The datasets generated and/or analysed during the current study are available in the dbGap repository: https://www.ncbi.nlm.nih.gov/projects/gap/cgi-bin/study.cgi?study_id=phs000007.v29.p10

**Competing interests.** The authors declare that they have no competing interests.

**Funding.** This project was supported in part by funding from the National Institutes of Health (NIH), including contracts N01-HC-25195 (RSV, MGL) and HHSN268201500001I (RSV, MGL), as well as NIH grants T32-ES007142 (JA), R01-HL134168 (SC, MJ), R01-ES027595 (MJ, SC), K01-HL135342 (OVD), and R01-HL134811 (OVD), the American Heart Association CVGPS Pathway Award (MGL, SC, MJ), Doris Duke Charitable Foundation Grant #2015092 (MJ, SC), the Tobacco Related Disease Research Program (MJ, JDW), and the Frontiers of Innovation Scholars Program (KAL).

**Authors' contributions.** BLC, JA, MH, MJ, and SC designed the study. BLC, JA, MH, JDW, KAL, GM, and AC conducted the experiments. BLC, JA, MGL, MJ, and SC analysed the data. JDW, KAL, OVD, RSV, and MGL provided material, data or analysis tools. BLC, JA, SJ, OVD, MJ, and SC wrote the paper. All authors read the paper and contributed to its final form.

**Acknowledgements.** Not applicable.

**Table 1. Data Structures Used for Analyses.**

| Dataset | Outcome Characteristics | No. of Metabolites | No. Observations (i.e. No. Persons) |
|---|---|---|---|
| 1 | Continuous | 200 | 200 |
| 2 | Continuous | 200 | 1000 |
| 3 | Continuous | 200 | 5000 |
| 4 | Continuous | 2000 | 200 |
| 5 | Continuous | 2000 | 1000 |
| 6 | Continuous | 2000 | 5000 |
| 7 | Binary: 20% frequency | 200 | 200 |
| 8 | Binary: 50% frequency | 200 | 1000 |
| 9 | Binary: 50% frequency | 200 | 5000 |
| 10 | Binary: 20% frequency | 2000 | 200 |
| 11 | Binary: 50% frequency | 2000 | 1000 |
| 12 | Binary: 50% frequency | 2000 | 5000 |

**Figure 1. Structure of the simulated dataset.** To perform statistical analyses within a controlled environment with pre-specified metabolite-outcome associations, we created a simulated dataset based generally on data features observed in multiple real-world datasets. One such simulated dataset demonstrates a scenario with multiple clusters of metabolites that have within-cluster correlation but little cross-cluster, mimicking the inter-relationships observed in actual experimentally derived human metabolomics studies.

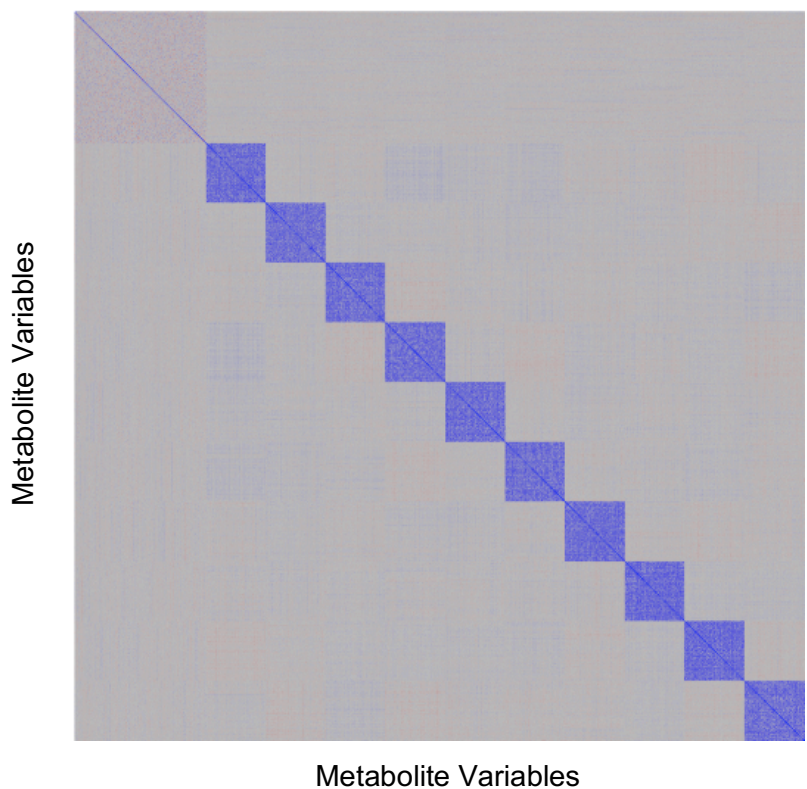

**Figure 2. Results for a continuous outcome.** The sensitivity, specificity, positive predictive value (PPV), negative predictive value (PPV), and false positive rate are displayed (as percent color fill of each bar) for each statistical method, reflecting their ability to correctly identify the top ten simulated metabolite associations, across varying numbers of total metabolite measures (M=200, or M=2000) in study samples collected from varying numbers of study subjects (N=200, N=1000, or N=5000). PCR, principal components regression; BON, Bonferroni; FDR, false discovery rate; LASSO, lease absolute shrinkage and selection operator; SPLS, sparse partial least squares; RF, random forests.

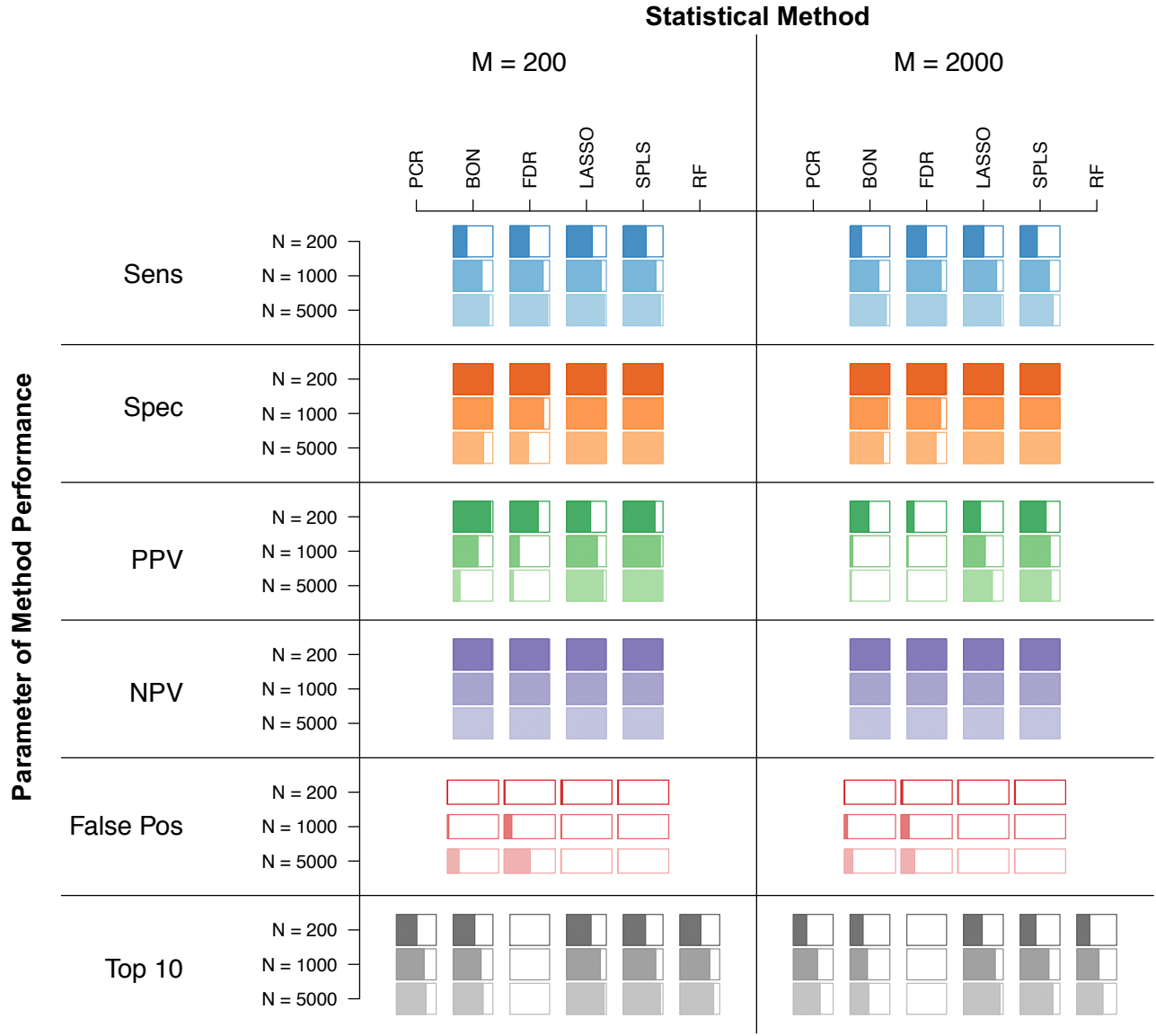

**Figure 3. Results for a binary outcome.** The sensitivity, specificity, positive predictive value (PPV), negative predictive value (PPV), and false positive rate are displayed (as percent color fill of each bar) for each statistical method, reflecting their ability to correctly identify the top ten simulated metabolite associations, across varying numbers of total metabolite measures (M=200, or M=2000) in study samples collected from varying numbers of study subjects (N=200, N=1000, or N=5000). PCR, principal components regression; BON, Bonferroni; FDR, false discovery rate; LASSO, lease absolute shrinkage and selection operator; SPLS, sparse partial least squares; RF, random forests.

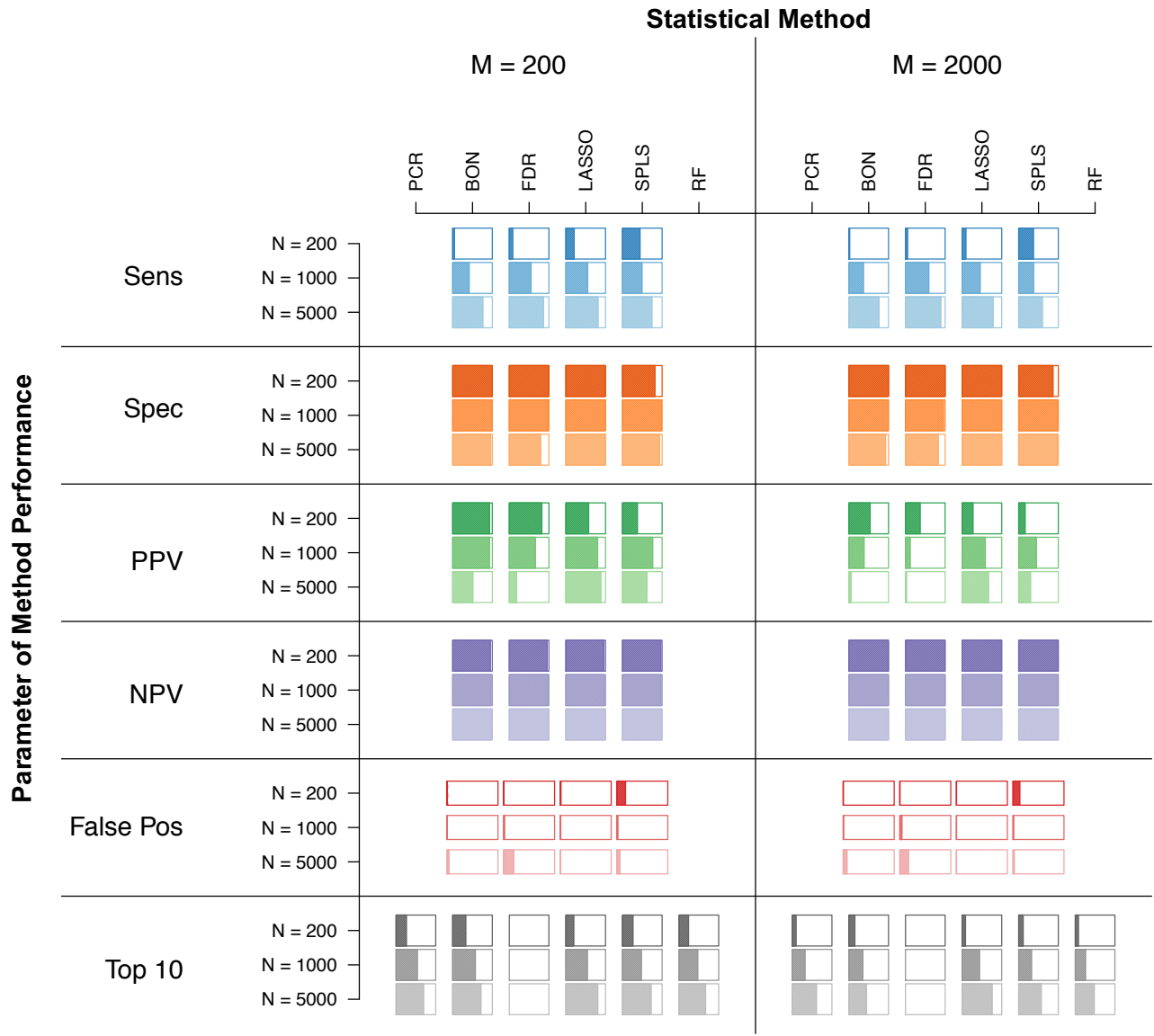

**Figure 4. Analysis of actual, experimentally derived metabolomics data.** The number of metabolites found in association with age (continuous outcome) and sex (binary outcome) from experimentally derived metabolomics studies (see text) for different statistical methods applied: false discovery rate (FDR), sparse partial least squares (SPLS), and least absolute shrinkage and selection operator (LASSO). The number of metabolite correlates found in common by the different methods is relatively small compared to the total number of apparently significantly associated metabolites.

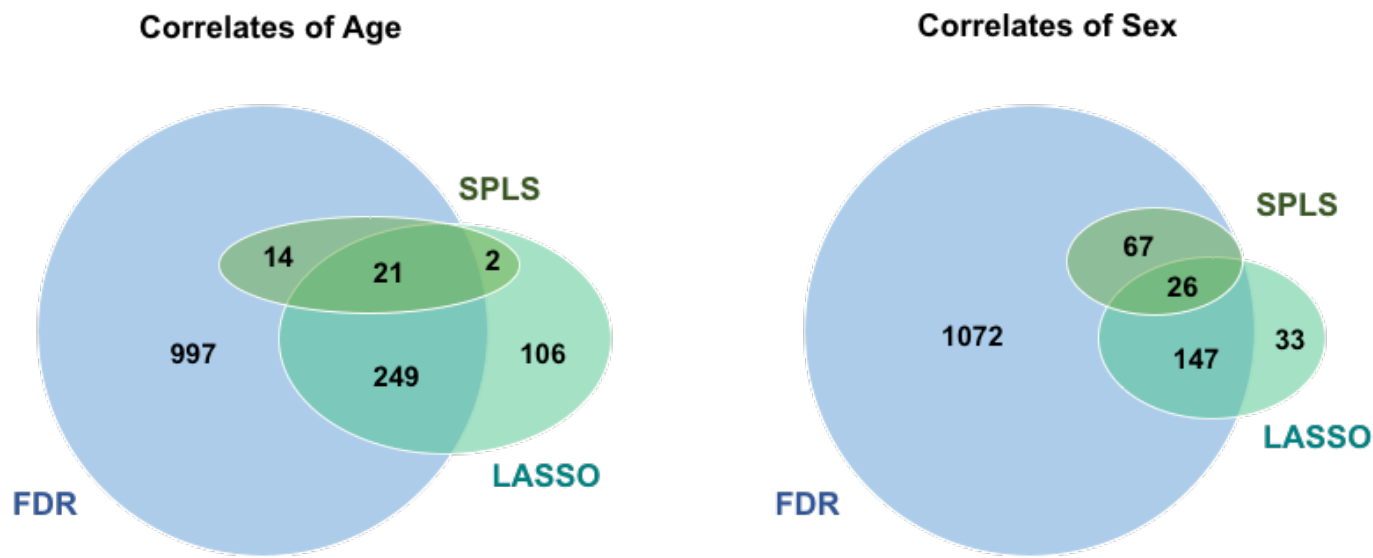

**Figure 5. Putative network distribution of metabolites identified by different methods used to analyze cohort-based metabolomics data.** The number of metabolites found in association with age (continuous outcome) and sex (binary outcome) from experimentally derived metabolomics studies (see text) for the different statistical methods applied was greater for traditional than for statistical learning models. Notably, the former identified metabolites that tended to be highly correlated with each other (Spearman rho ≥0.65), whereas the latter identified a more parsimonious number of metabolites distributed across the putative network of all highly inter-correlated metabolites. BON, Bonferroni; FDR, false discovery rate; LASSO, lease absolute shrinkage and selection operator; SPLS, sparse partial least squares.

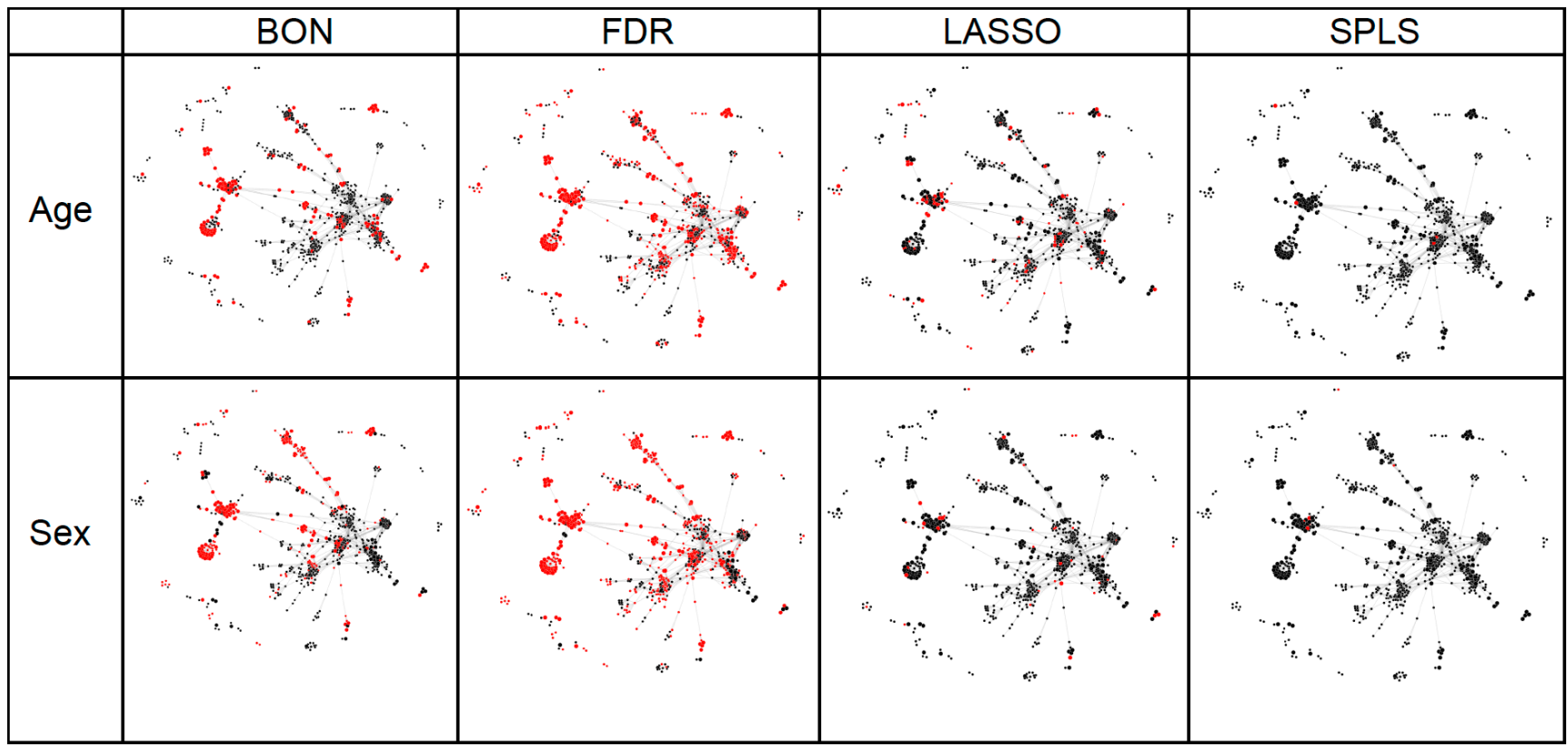

# SUPPLEMENTAL MATERIAL

**Table S1. Known Metabolite Compounds Assayed by the LC-MS Method Used**

| Metabolite Type | Mass-to-Charge Ratio | ID |
| --- | --- | --- |
| polar molecule | 227.0645 | DEOXYURIDINE |
| free fatty acid | 227.2012 | Myristic Acid |
| polar molecule | 241.0738 | LUMICHROME |
| free fatty acid | 241.2170 | Pentadecanoic Acid |
| polar molecule | 242.0801 | CYTIDINE |
| free fatty acid | 253.2172 | Palmitoleic Acid |
| free fatty acid | 255.2329 | Palmitic Acid |
| eicosanoid | 265.1812 | tetranor 12(R) HETE |
| polar molecule | 267.0475 | HOMOCYSTINE |
| free fatty acid | 267.2330 | Heptadecaenoic Acid |
| steroid | 269.1758 | Estrone |
| free fatty acid | 275.2020 | Stearidonic Acid |
| free fatty acid | 277.2175 | GAMMA-LINOLENATE |
| free fatty acid | 279.2331 | LINOLEATE |
| free fatty acid | 281.2489 | ELAIDATE |
| eicosanoid | 291.1967 | 13S-HpOTrE(gamma) |
| eicosanoid | 293.2122 | 13-oxoODE |
| eicosanoid | 293.2122 | 9-oxoODE |
| eicosanoid | 295.2275 | 9-HODE |
| eicosanoid | 299.2010 | 15-oxoETE |
| eicosanoid | 299.2039 | 5-oxoETE |
| free fatty acid | 301.2172 | Eicosapentaenoic Acid |
| free fatty acid | 303.2331 | Arachidonic Acid |
| free fatty acid | 305.2486 | Eicosatrienoic Acid |
| free fatty acid | 307.2644 | Eicosadienoic Acid |
| free fatty acid | 309.2044 | MYRISTATE |
| free fatty acid | 309.2798 | Gondolic Acid |
| free fatty acid | 311.2955 | ARACHIDATE |
| free fatty acid | 313.2387 | PALMITOLEATE |

| | | | |
|---|---|---|---|
| eicosanoid | 315.1951 | | 15-keto-PGA1 |
| eicosanoid | 315.1964 | | 13(S) HOTrE(y) |
| eicosanoid | 315.1971 | | bicyclo PGE2 |
| eicosanoid | 315.2000 | | 5S-HpEPE |
| eicosanoid | 315.2000 | | 15d PGJ2 |
| eicosanoid | 317.2110 | | HXA3 |
| eicosanoid | 317.2115 | | 5(S) HEPE |
| eicosanoid | 317.2117 | | 18(S) HEPE |
| eicosanoid | 317.2118 | | 15(S) HEPE |
| eicosanoid | 317.2121 | | 14(15) EpETE |
| eicosanoid | 317.2128 | | 12epi LTB4 |
| eicosanoid | 317.2136 | | 5,15-diHETE |
| eicosanoid | 319.2260 | | 16-HETE |
| eicosanoid | 319.2278 | | 11-HETE |
| eicosanoid | 319.2283 | | 14,15-EET |
| eicosanoid | 319.2291 | | 5,6-EET |
| eicosanoid | 321.1712 | | 11-dehydro-2,3-dinor-TXB2 |
| eicosanoid | 321.2423 | | 8(S) HETrE |
| eicosanoid | 321.2435 | | 15(S) HETrE |
| polar molecule | 323.0974 | | CELLOBIOSE |
| endocannabinoid | 326.3038 | | Stearoyl EA |
| free fatty acid | 327.2326 | | Docosahexaenoic Acid |
| polar molecule | 329.0161 | | DEOXYURIDINE-MONOPHOSPHATE |
| eicosanoid | 331.1890 | | 9S-HpOTrE |
| steroid | 331.1909 | | Estradiol |
| eicosanoid | 331.1916 | | PGD3 |
| eicosanoid | 333.2061 | | 8-iso-15-keto-PGF2alpha |
| eicosanoid | 333.2064 | | PGA2 |
| eicosanoid | 333.2070 | | 12oxo LTB4 |
| eicosanoid | 333.2070 | | dhk PGE2 |
| eicosanoid | 333.2071 | | 13,14-dihydro-15-keto-PGA2 |
| eicosanoid | 333.2072 | | ent-PGE2 |
| eicosanoid | 333.2074 | | LXB4 |
| eicosanoid | 333.2077 | | 20cooh AA |
| eicosanoid | 333.2080 | | 8-iso-PGA2 |
| free fatty acid | 333.2811 | | Docosatrienoic Acid |
| eicosanoid | 335.2222 | | 15S-HpETE |
| eicosanoid | 335.2226 | | PGF2beta |
| eicosanoid | 335.2228 | | 8,12-iso-iPF2Ã-VI-1,5-lactone |
| eicosanoid | 335.2228 | | 14,15-DiHETE |
| eicosanoid | 335.2228 | | 8-iso-PGA1 |
| eicosanoid | 335.2232 | | 15R-PGE1 |
| eicosanoid | 335.2234 | | 15R-PGF2alpha |
| eicosanoid | 335.2249 | | 5,6-diHETE |
| free fatty acid | 335.2959 | | Docosadienoic Acid |
| free fatty acid | 337.3116 | | ERUCATE |
| eicosanoid | 339.2175 | | 12-HHTrE |
| free fatty acid | 339.2533 | | Linoleic Acid |
| free fatty acid | 339.3269 | | Behenic Acid |
| eicosanoid | 341.2163 | | 17S-HpDHA |
| free fatty acid | 343.2853 | | Stearic Acid |
| eicosanoid | 351.2170 | | 9-oxoOTrE |
| eicosanoid | 351.2202 | | 8-iso-PGE2 |
| free fatty acid | 351.3270 | | Tricosenoic Acid |
| eicosanoid | 353.2319 | | 13(S) HOTrE |
| eicosanoid | 353.2337 | | 9(S) HOTrE |
| eicosanoid | 353.2343 | | 11b dhk PGF2a |
| free fatty acid | 353.3427 | | TRICOSANOATE |
| eicosanoid | 355.2486 | | 13-HODE |
| eicosanoid | 355.2486 | | 12,13 EpOME |
| endocannabinoid | 358.2962 | | Palmitoyl Ethanolamide |
| endocannabinoid | 358.2966 | | Palmitoyl EA or |
| steroid | 359.1867 | | Cortisone |
| eicosanoid | 359.2222 | | 5,6-diHETrE |
| free fatty acid | 359.2971 | | Docosaenoic Acid |
| eicosanoid | 363.2556 | | 1a,1b-dihomo-PGE1 |
| eicosanoid | 363.2557 | | dihomo PGF2a |
| polar molecule | 364.0595 | | CYTIDINE 2',3'-CYCLIC PHOSPHATE |
| free fatty acid | 365.2658 | | STEARATE |
| free fatty acid | 367.3583 | | Lignoceric Acid |
| bile acid | 375.2906 | | Lithocholic Acid |
| eicosanoid | 379.2486 | | 15-HETE |
| endocannabinoid | 382.2975 | | Linoleoyl EA |
| endocannabinoid | 384.3114 | | Oleoyl Ethanolamide |
| endocannabinoid | 384.3128 | | Oleoyl EA |
| free fatty acid | 387.3286 | | Nervonic Acid |
| steroid | 389.2309 | | 11-deoxycortisosterone or 17a-hydroxyprogesterone |
| bile acid | 389.2697 | | Cholic Acid |
| free fatty acid | 391.2850 | | Adrenic Acid |
| bile acid | 391.2862 | | Deoxycholic Acid |
| bile acid | 391.2863 | | Ursodeoxycholic acid |
| eicosanoid | 393.2280 | | 15d PGD2 |
| eicosanoid | 393.2294 | | PGB2 |
| eicosanoid | 395.2429 | | LTB4 |
| eicosanoid | 395.2433 | | 15-epi-PGA1 |
| eicosanoid | 395.2442 | | PGA1 |
| eicosanoid | 395.2445 | | 12,13 diHOME |
| polar molecule | 397.3365 | | ERUCATE |
| polar molecule | 401.1297 | | PALATINOSE |
| eicosanoid | 403.2485 | | 15 oxoEDE |
| eicosanoid | 403.2500 | | 14 HDoHE |
| steroid | 405.2265 | | Unk |
| steroid | 405.2271 | | Unk |
| steroid | 405.2277 | | 11-deoxycortisol |
| steroid | 405.2287 | | Unk |
| steroid | 405.2295 | | CORTEXOLONE |
| polar molecule | 407.0623 | | INOSINE MONOPHOSPHATE |
| bile acid | 407.2804 | | b-Muricholic Acid |
| eicosanoid | 411.2365 | | 6S-LXA4 |
| endocannabinoid | 415.3076 | | OLEOYL-GLYCEROL |
| eicosanoid | 417.2246 | | 12S-HpETE |

| | | |
|---|---|---|
| eicosanoid | 417.2284 | 8,15-diHETE |
| steroid | 419.2078 | CORTISONE |
| steroid | 421.2239 | CORTISOL |
| steroid | 421.2241 | Cortisol |
| endocannabinoid | 423.3101 | 2-AG ether |
| steroid | 425.2547 | allo-Tetrahydrocortisol |
| free fatty acid | 425.3617 | NERVONATE |
| bile acid | 430.2965 | Glycodeoxycholic Acid |
| endocannabinoid | 430.3028 | Docosahexanoyl EA |
| bile acid | 435.3122 | LITHOCHOLATE |
| endocannabinoid | 437.2899 | 2-AG maybe |
| bile acid | 448.3074 | Glycoursodeoxycholic Acid |
| eicosanoid | 451.2348 | 6k PGF1a |
| bile acid | 451.3071 | Chenodeoxycholic acid |
| bile acid | 464.3021 | GLYCOCHOLATE |
| bile acid | 470.2848 | Glycochenodeoxycholic Acid |
| bile acid | 480.2745 | Tauroursodeoxycholic acid |
| eicosanoid | 495.2607 | 14,15 LTD4 |
| bile acid | 498.2904 | Taurodeoxycholic Acid |
| bile acid | 514.2843 | Taurocholic Acid |
| bile acid | 520.2656 | Taurochenodeoxycholic acid |
| polar molecule | 540.0464 | ADENOSINE DIPHOSPHATE RIBOSE |
| polar molecule | 610.0507 | ADP-GLUCOSE |

**Figure S1. Estimates of power to detect metabolites associated with a continuous outcome for a given effect size based on simulations.** The estimated power to detect the top ten metabolite associations per effect size is shown for each statistical method, across varying numbers of total metabolite measures (M=200, or M=2000) in study samples collected from varying numbers of study subjects (N=200, N=1000, or N=5000). BON, Bonferroni; FDR, false discovery rate; LASSO, lease absolute shrinkage and selection operator; SPLS, sparse partial least squares.

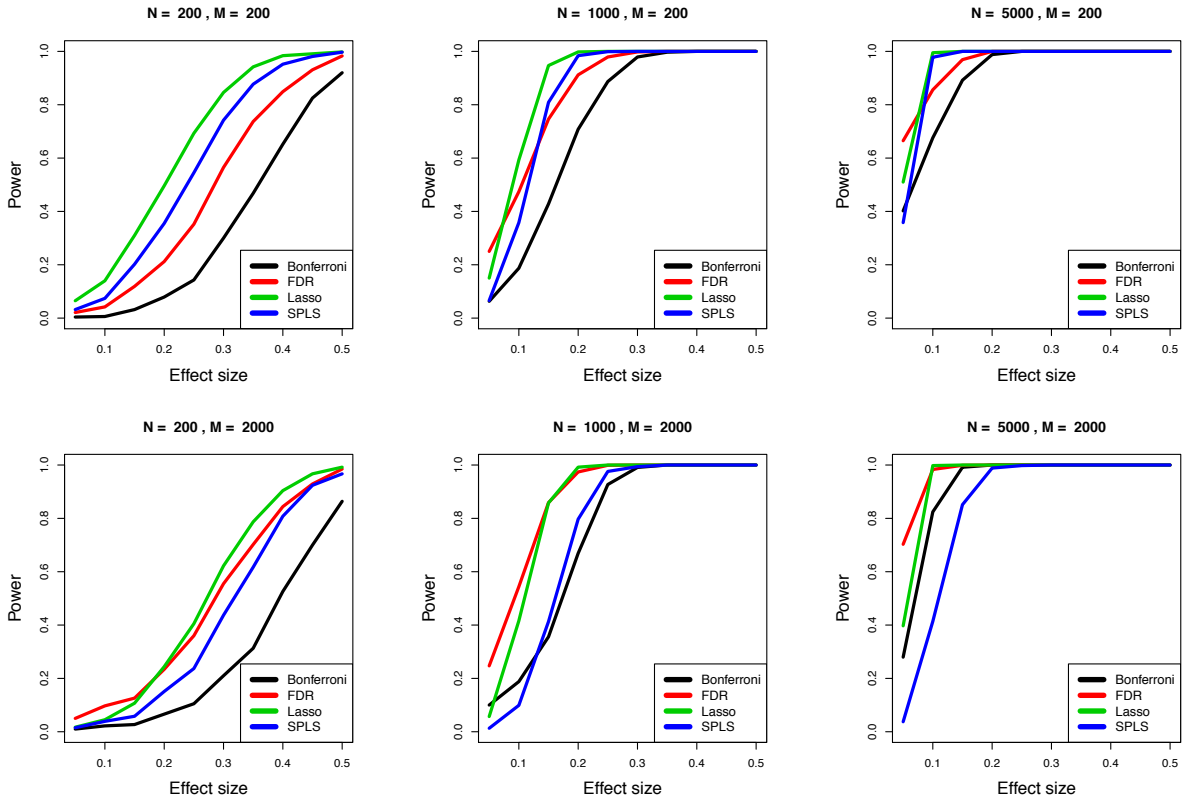

**Figure S2. Estimates of power to detect metabolites associated with a binary outcome for a given effect size based on simulations.** The estimated power to detect the top ten metabolite associations per effect size is shown for each statistical method, across varying numbers of total metabolite measures (M=200, or M=2000) in study samples collected from varying numbers of study subjects (N=200, N=1000, or N=5000). BON, Bonferroni; FDR, false discovery rate; LASSO, lease absolute shrinkage and selection operator; SPLS, sparse partial least squares.

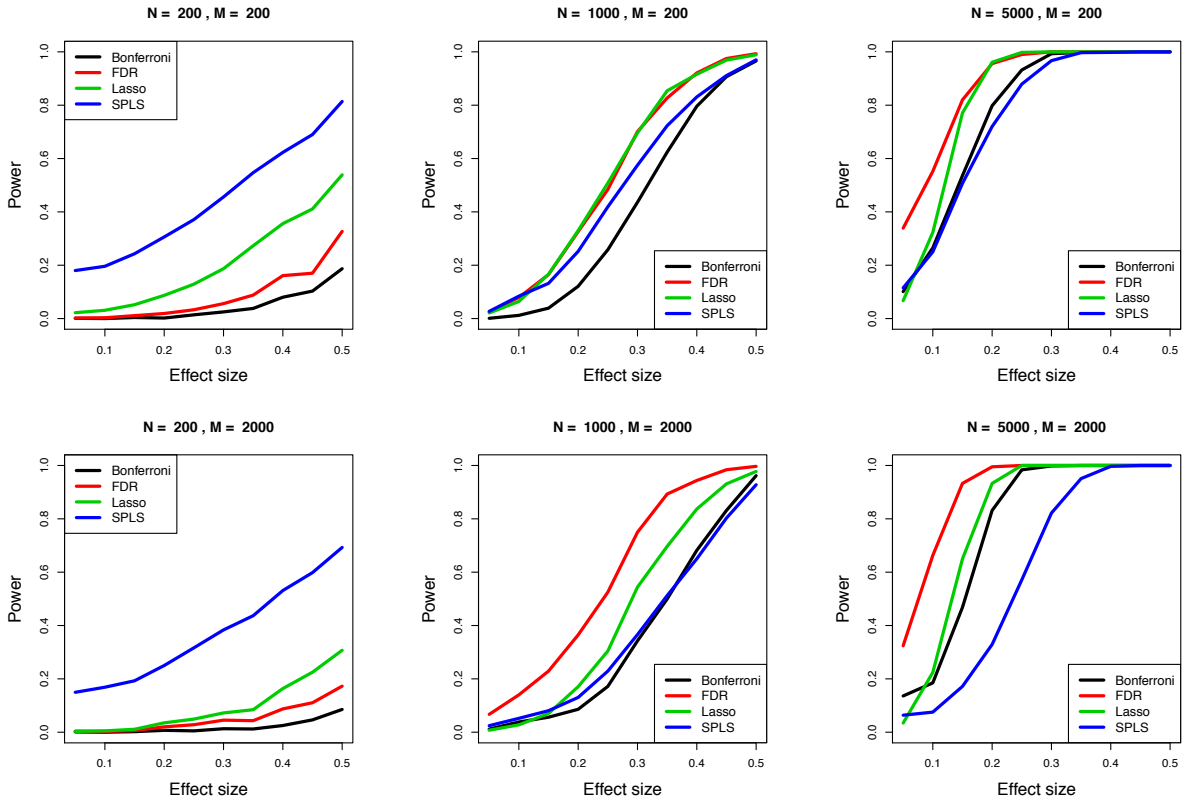

**Figure S3. Results for a continuous outcome based on simulations with positive and negative inter-metabolite correlations.** The sensitivity, specificity, positive predictive value (PPV), negative predictive value (PPV), and false positive rate are displayed (as percent color fill of each bar) for each statistical method, reflecting their ability to correctly identify the top ten simulated metabolite associations, across varying numbers of total metabolite measures (M=200, or M=2000) in study samples collected from varying numbers of study subjects (N=200, N=1000, or N=5000). PCR, principal components regression; BON, Bonferroni; FDR, false discovery rate; LASSO, lease absolute shrinkage and selection operator; SPLS, sparse partial least squares; RF, random forests.

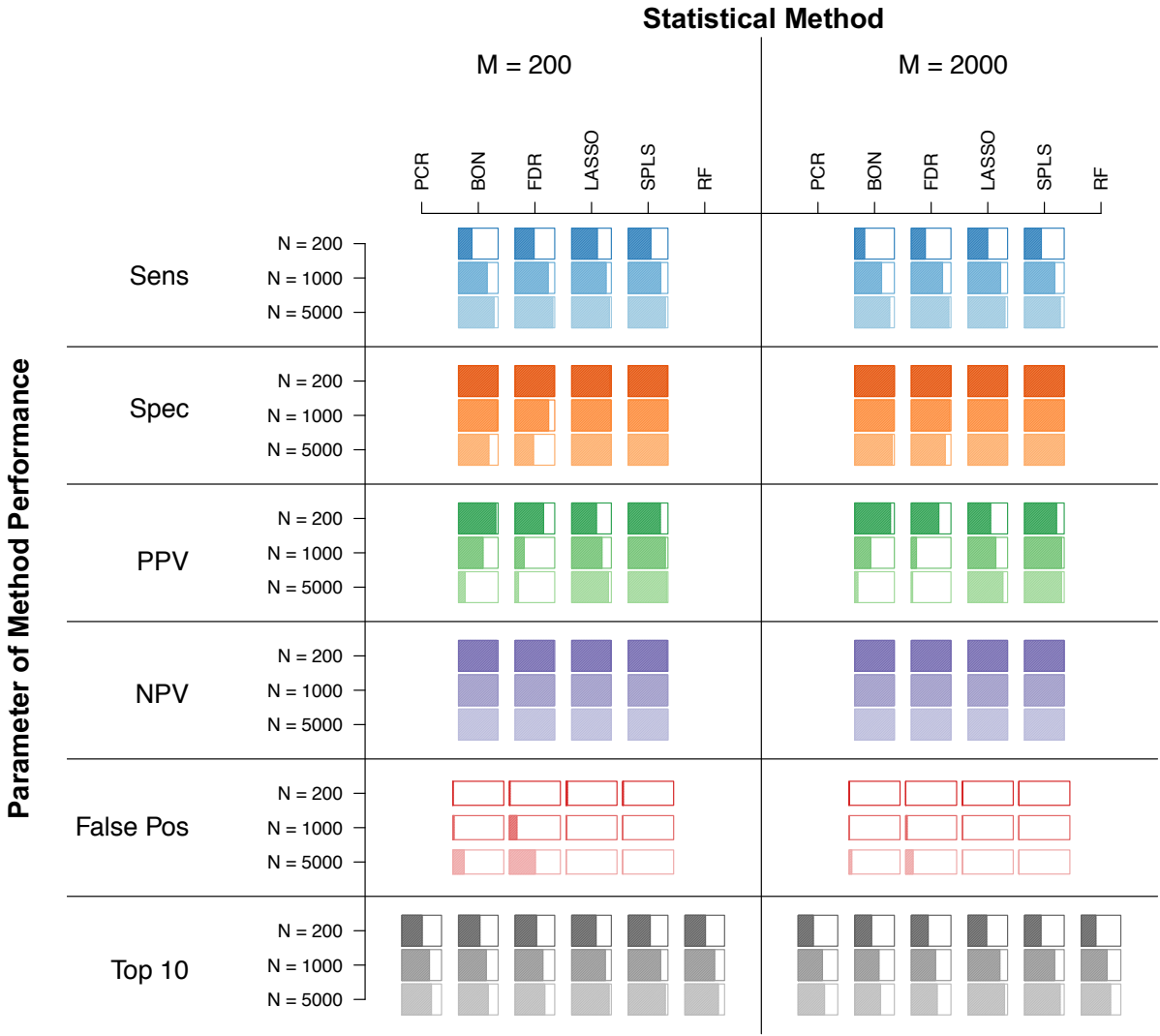

**Figure S4. Results for a binary outcome based on simulations with positive and negative inter-metabolite correlations.** The sensitivity, specificity, positive predictive value (PPV), negative predictive value (PPV), and false positive rate are displayed (as percent color fill of each bar) for each statistical method, reflecting their ability to correctly identify the top ten simulated metabolite associations, across varying numbers of total metabolite measures (M=200, or M=2000) in study samples collected from varying numbers of study subjects (N=200, N=1000, or N=5000). PCR, principal components regression; BON, Bonferroni; FDR, false discovery rate; LASSO, lease absolute shrinkage and selection operator; SPLS, sparse partial least squares; RF, random forests.

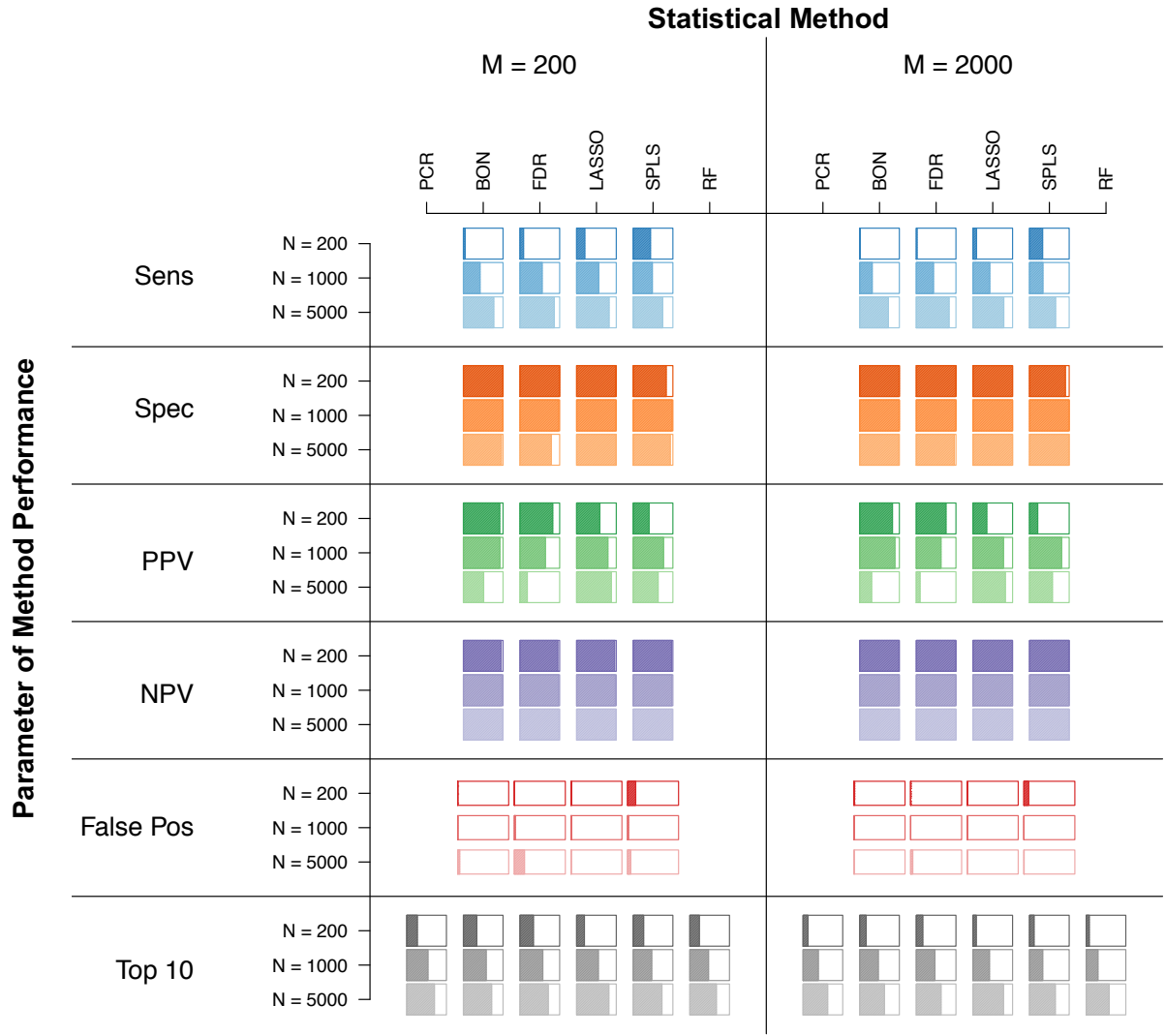

**Figure S5. Results for a continuous outcome based on simulations with highly correlated important covariates.** The sensitivity, specificity, positive predictive value (PPV), negative predictive value (PPV), and false positive rate are displayed (as percent color fill of each bar) for each statistical method, reflecting their ability to correctly identify the top ten simulated metabolite associations, across varying numbers of total metabolite measures (M=200, or M=2000) in study samples collected from varying numbers of study subjects (N=200, N=1000, or N=5000). PCR, principal components regression; BON, Bonferroni; FDR, false discovery rate; LASSO, lease absolute shrinkage and selection operator; SPLS, sparse partial least squares; RF, random forests.

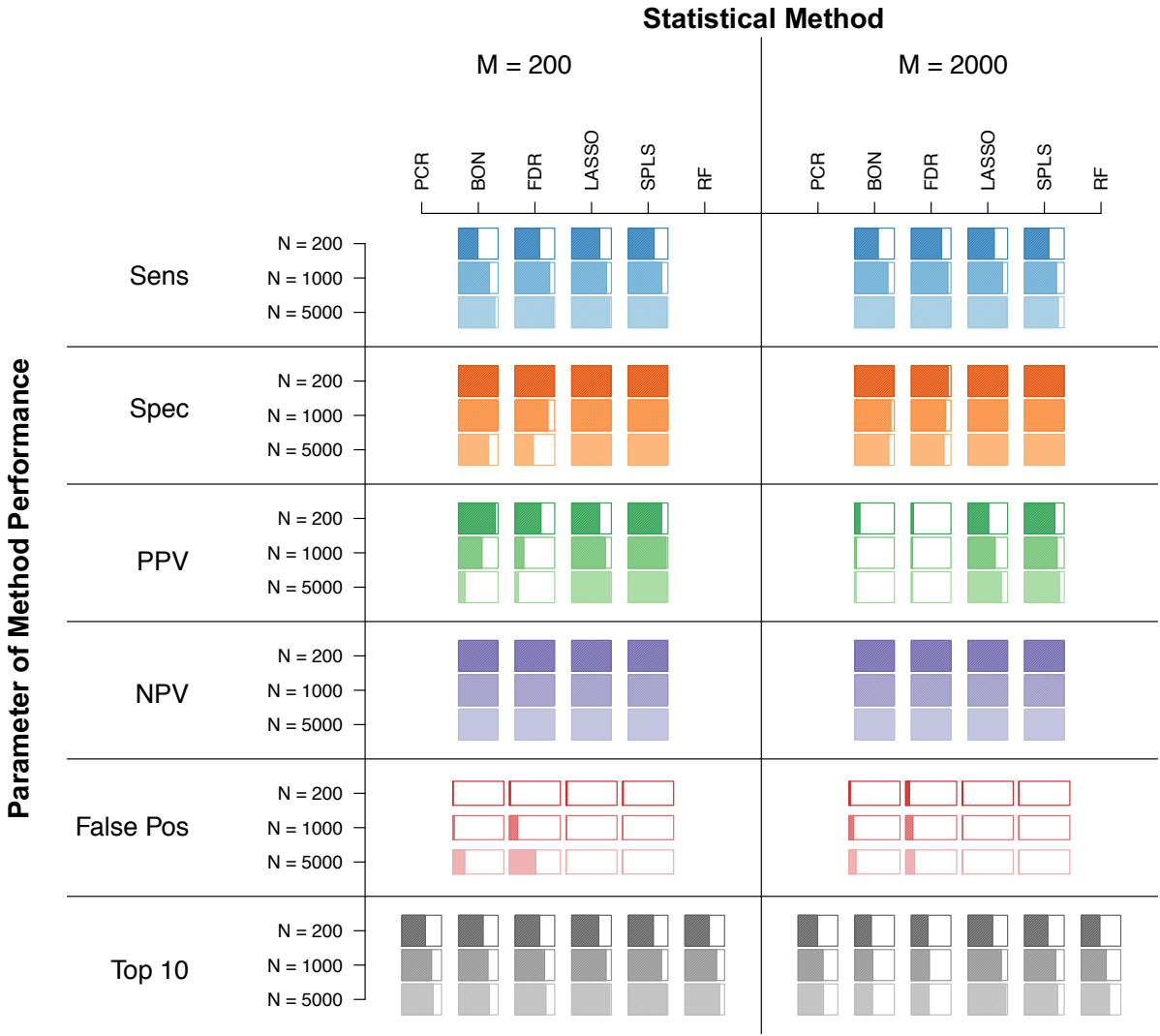

**Figure S6. Results for a binary outcome based on simulations with highly correlated important covariates.** The sensitivity, specificity, positive predictive value (PPV), negative predictive value (PPV), and false positive rate are displayed (as percent color fill of each bar) for each statistical method, reflecting their ability to correctly identify the top ten simulated metabolite associations, across varying numbers of total metabolite measures (M=200, or M=2000) in study samples collected from varying numbers of study subjects (N=200, N=1000, or N=5000). PCR, principal components regression; BON, Bonferroni; FDR, false discovery rate; LASSO, lease absolute shrinkage and selection operator; SPLS, sparse partial least squares; RF, random forests.

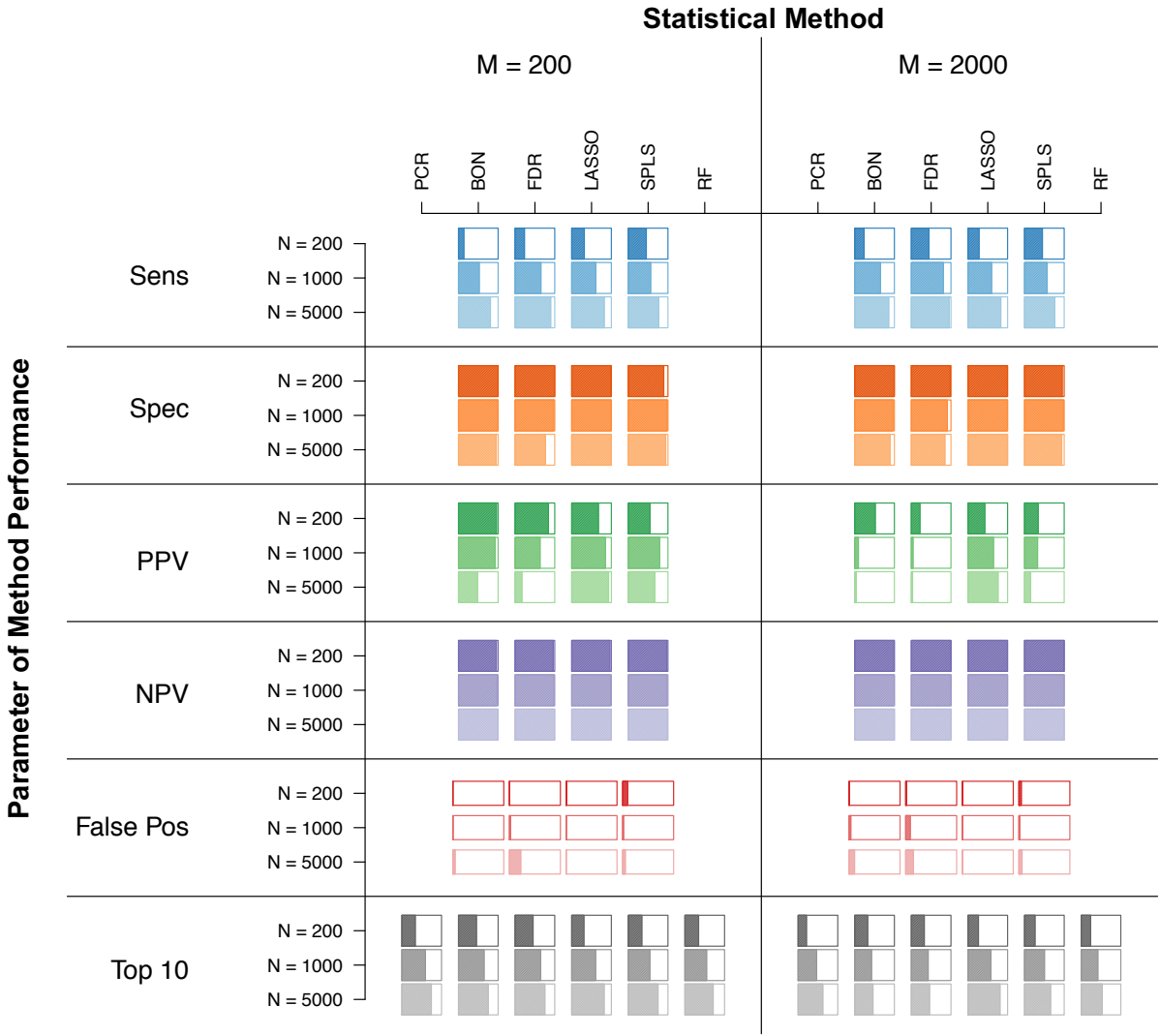

**Figure S7. Results of analyzing actual, experimentally derived metabolomics data while controlling for batch effects.** The number of metabolites found in association with age (continuous outcome) and sex (binary outcome) from experimentally derived metabolomics studies (see text) for different statistical methods applied: false discovery rate (FDR), sparse partial least squares (SPLS), and least absolute shrinkage and selection operator (LASSO). The number of metabolite correlates found in common by the different methods is relatively small compared to the total number of apparently significantly associated metabolites.

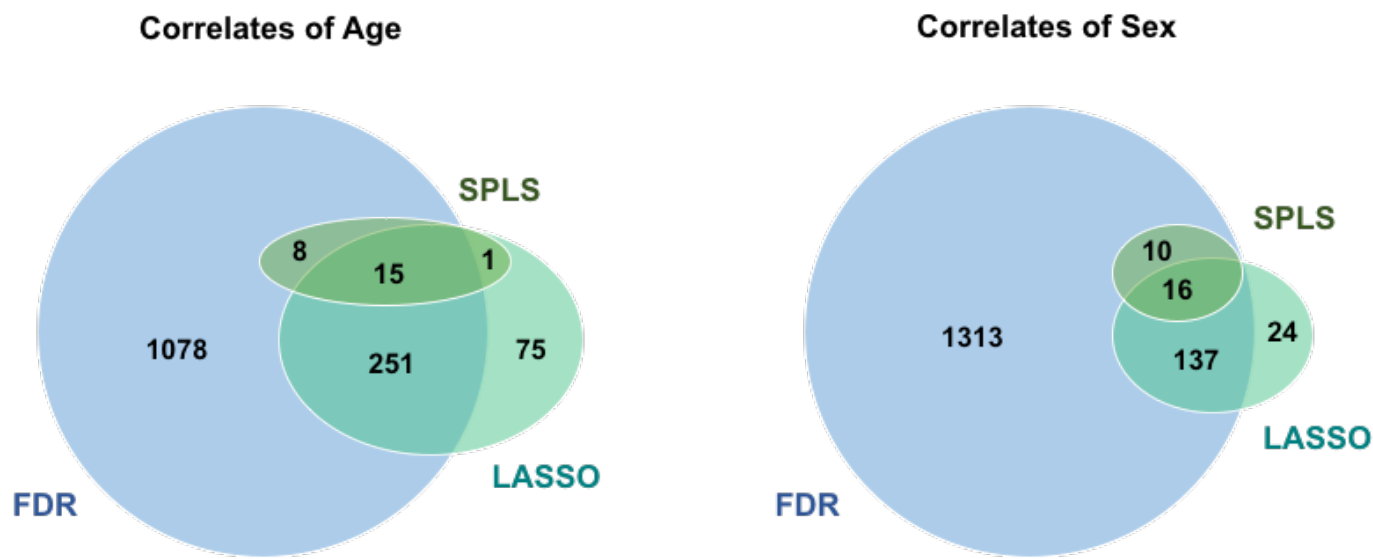